\documentclass[pra,twocolumn,showpacs,amsmath,amssymb]{revtex4-1}

\usepackage[T1]{fontenc} 
\usepackage{bm}
\usepackage[bookmarks=false]{hyperref}

\newcommand{\bra}[1]{\langle #1|}
\newcommand{\ket}[1]{|#1\rangle}
\newcommand{\braket}[1]{\langle #1 \rangle}

\def\dd{\mathrm{d}}
\def\ee{\mathrm{e}}
\def\ii{\mathrm{i}}
\def\vnabla{\bm{\nabla}}

\def\div{\vnabla\cdot}
\def\rot{\vnabla\times}

\def\const{\mathrm{const.}}

\def\mdie{\bm{\varepsilon}}
\def\mchi{\bm{\chi}}
\def\diez{\varepsilon_0}
\def\muz{\mu_0}
\def\vol{V}

\def\intD{\int_{\mathcal{D}}}

\def\dm{d}

\def\wa{\omega_{\text{a}}}
\def\wc{\omega_{\text{c}}}
\def\rabi{g}
\def\rabid{\tilde{g}}

\def\oH{\hat{H}}
\def\oHC{\hat{H}_{\text{Coulomb}}}

\def\oHCO{\hat{H}_{\text{Coulomb}}^{\text{1D}}}
\def\oHP{\hat{H}_{\text{PZW}}}

\def\oHPO{\hat{H}_{\text{PZW}}^{\text{1D}}}
\def\oHrad{\hat{H}_{\text{rad}}}
\def\oHatom{\hat{H}_{\text{atom}}}

\def\oVC{\hat{V}_{\text{C}}}

\def\ovr{\hat{\bm{r}}}
\def\ovp{\hat{\bm{p}}}
\def\ovA{\hat{\bm{A}}}
\def\ovE{\hat{\bm{E}}}
\def\ovET{\hat{\bm{E}}_{\perp}}

\def\ovDT{\hat{\bm{D}}_{\perp}}
\def\ovPi{\hat{\bm{\varPi}}}
\def\ovP{\hat{\bm{P}}}
\def\ovPT{\hat{\bm{P}}_{\perp}}
\def\ovPL{\hat{\bm{P}}_{\parallel}}

\def\oa{\hat{a}}
\def\oad{\hat{a}^{\dagger}}
\def\ob{\hat{b}}
\def\obd{\hat{b}^{\dagger}}
\def\osigma{\hat{\sigma}}
\def\osigmad{\hat{\sigma}^{\dagger}}

\def\vunit{\bm{e}}
\def\vr{\bm{r}}
\def\vR{\bm{R}}
\def\vk{\bm{k}}

\def\vP{\bm{P}}
\def\vE{\bm{E}}

\def\munit{\bm{1}}
\def\mG{\bm{\mathsf{G}}}

\def\mdeltaL{\bm{\delta}^{\parallel}}
\def\mdeltaT{\bm{\delta}^{\perp}}

\begin{document}


\title{Stability of polarizable materials against superradiant phase transition}

\author{Motoaki Bamba}
\altaffiliation{E-mail: bamba@acty.phys.sci.osaka-u.ac.jp}
\affiliation{Department of Physics, Osaka University, 1-1 Machikaneyama, Toyonaka, Osaka 560-0043, Japan}
\author{Tetsuo Ogawa}
\affiliation{Department of Physics, Osaka University, 1-1 Machikaneyama, Toyonaka, Osaka 560-0043, Japan}
\date{\today}

\begin{abstract}
The possibility of the superradiant phase transition
in polarizable materials described by the minimal-coupling Hamiltonian
with the longitudinal dipole-dipole interaction is examined.
We try to reduce the Hamiltonian into the Dicke one
in homogeneous and infinite case,
and discuss the stability of normal ground state
by the formalism of Green function
in spatially inhomogeneous case.
The presence of the longitudinal dipole-dipole interaction
does not enable the superradiant phase transition,
if the transverse and longitudinal fields are decoupled.
Although the full dipole-dipole interaction can be eliminated
in the electric-dipole gauge in the absence of overlap between individual atomic dipoles,
we cannot reduce the Hamiltonian to the Dicke one,
because the elimination is justified
only if all the transverse and longitudinal fields remain.
Even if the transverse and longitudinal fields are mixed
in spatially inhomogeneous  systems,
the normal ground state is still stable
if the system does not show the superradiant phase transition
in the homogeneous case.
\end{abstract}



\maketitle
\section{Introduction}
The light-matter interaction has been discussed
by the classical electrodynamics,
in which the Maxwell equations describe
the dynamics of the electromagnetic fields in matters,
and also by the quantum electrodynamics \cite{cohen-tannoudji89}.
In the latter regime,
an ensemble of atoms interacting with the radiation field in a cavity
has been supposed as a standard model for discussing a variety of optical phenomena,
e.g., studying the laser \cite{Haken1970},
which are well described by the Dicke Hamiltonian
\begin{equation} \label{eq:oHDicke} 
\oH_{\text{Dicke}} = \hbar\wc\oad\oa
+ \sum_{\lambda=1}^N \frac{\hbar\wa}{2}\osigma_{\lambda}^z
+ \sum_{\lambda=1}^N \frac{\ii\hbar\rabid}{\sqrt{N}}(\oad\osigma_{\lambda}-\osigmad_{\lambda}\oa).
\end{equation}
Here, $\oa$ is the annihilation operator of a photon in a cavity mode with a frequency $\wc$.
$\osigma_{\lambda}^z = \ket{e_{\lambda}}\bra{e_{\lambda}}-\ket{g_{\lambda}}\bra{g_{\lambda}}$ and
$\osigma_{\lambda} = \ket{g_{\lambda}}\bra{e_{\lambda}}$
are operators involving the $\lambda$-th atom
with ground state $\ket{g_{\lambda}}$ and excited one $\ket{e_{\lambda}}$,
and $\wa$ is the frequency difference between them.
$N$ is the number of atoms, and $\rabid$ represents the strength of the light-matter coupling.
When the coupling strength goes into the ultrastrong regime
($\rabid\gtrsim\wa, \wc$),
it is known \cite{Mallory1969PR,Hepp1973AP,Wang1973PRA}
that the radiation field and the atomic polarization get non-zero amplitudes
even in the ground state of the system described by the Dicke Hamiltonian \eqref{eq:oHDicke}.
This is called the superradiant phase transition (SPT).
However, it is also known
that the SPT is prevented by the presence of the $A^2$ term
\cite{Rzazewski1975PRL,Rzazewski1976PRA,Yamanoi1976PLA,Yamanoi1979JPA}
\begin{equation} \label{eq:oHA2} 
\oH_{A^2} = \frac{\hbar\rabid^2}{\wa}(\oa+\oad)(\oa+\oad)
\end{equation}
appearing in the Coulomb gauge \cite{cohen-tannoudji89},
and also by the $P^2$ term 
\begin{equation} \label{eq:oHP2} 
\oH_{P^2} = \sum_{\lambda,\lambda'}\frac{\hbar\rabid^2}{N\wc}
(\osigma_{\lambda}+\osigma^{\dagger}_{\lambda})(\osigma_{\lambda'}+\osigma^{\dagger}_{\lambda'})
\end{equation}
appearing in the electric-dipole gauge \cite{Emeljanov1976PLA,Yamanoi1978P2}.
Further, the SPT was also discussed from the viewpoint of gauge-invariance
\cite{Woolley1976JPA,Knight1978PRA,Bialynicki-Birula1979PRA}.
Concerning the minimal-coupling Hamiltonian
with the Coulomb interaction
[in the Coulomb gauge and shown in Eq.~\eqref{eq:oHC}],
the SPT is forbidden at least under the dipole approximation
(no-go theorem) \cite{Bialynicki-Birula1979PRA,Rzazewski1991PRA},
and is also denied beyond that approximation
but in the classical treatment of the radiation field
\cite{Gawedzki1981PRA}.
Due to the gauge-invariance, the SPT in the same system
is denied also in the electric-dipole gauge.

Although we do not yet find any real systems for obtaining the SPT
in equilibrium situations,
it is recently recognized that the SPT can occur effectively
at least in non-equilibrium situations (driven-dissipative systems).
It was suggested theoretically that an analog of the Dicke Hamiltonian
(without $A^2$ or $P^2$ term) can be constructed in a system of multi-level atoms
under coherent pumping \cite{Dimer2007PRA},
and a phase transition corresponding to the SPT has been observed
experimentally in the system of cold atoms \cite{Baumann2010N}.
Further, in recent years,
the ultrastrong light-matter coupling has also been realized in a variety of systems
\cite{Gunter2009N,Anappara2009PRB,Todorov2009PRL,Todorov2010PRL,
Niemczyk2010NP,Fedorov2010PRL,Forn-Diaz2010PRL,
Schwartz2011PRL,Porer2012PRB,Scalari2012S}.
Now, the possibility of the SPT re-attracts an attention
\cite{Nataf2010NC,Viehmann2011PRL,Hagenmuller2012PRL,Chirolli2012PRL}
in the direction of how the Dicke Hamiltonian is constructed
and the SPT occurs in real systems,
although it is not yet observed in the equilibrium situations.

In order to investigate the systems with the ultrastrong light-matter coupling
quantitatively and qualitatively,
it is inevitable to understand precisely the Hamiltonian of those systems,
i.e., whether the $A^2$ or $P^2$ term exists
and the SPT can occur even in the equilibrium situations.
For example, its possibility involving the magnetic-dipole interaction
involving electron spins \cite{Knight1978PRA} is not yet denied.
The gauge transformation under finite-level and finite-mode approximations,
which are performed for deriving the Dicke Hamiltonian \eqref{eq:oHDicke}
and destroy the gauge invariance,
is also recognized as a problem from the experimental viewpoint
in the ultrastrong light-matter coupling regime \cite{Todorov2010PRL,Todorov2012PRB,Todorov2014PRB}.

Further, it was recently
pointed out in Refs.~\cite{Keeling2007JPCM,Vukics2014PRL}
that the $P^2$ term seems to disappear in the electric-dipole gauge
thanks to the presence of the longitudinal dipole-dipole interaction
and the absence of overlap between individual atomic dipoles
(atoms are well separated from each other),
and the SPT seems to occur at first glance.
This indication does not violate the gauge invariance,
and avoid the no-go theorem \cite{Bialynicki-Birula1979PRA,Rzazewski1991PRA}
by the modification of the Hamiltonian,
although the model itself seems to be included in those discussed by the no-go theorem
\cite{Bialynicki-Birula1979PRA,Rzazewski1991PRA}
in the Coulomb gauge under the dipole approximation.
In this paper, we discuss whether this elimination of the $P^2$ term
is really justified for the system consisting of atoms with radiative transition dipoles
(polarizable atoms)
not only in spatially homogeneous and infinite case but also in inhomogeneous (finite) case.
In the latter case, we examine the stability of normal ground state
(with no field amplitude) in the inhomogeneous systems
according to the formalism of Green function of the electromagnetic fields
in matters \cite{knoll01,abrikosov75ch6}.
From these discussions, we can understand correctly
the Hamiltonians under the finite-level and finite-mode approximations,
which are desired for investigating realistic systems with the ultrastrong light-matter coupling.

We first review the standard derivation of the Dicke-like Hamiltonians
in the Coulomb and electric-dipole gauges in Sec.~\ref{sec:standard}.
The elimination of the $P^2$ term proposed in Refs.~\cite{Keeling2007JPCM,Vukics2014PRL}
and its validity in homogeneous systems are discussed in Sec.~\ref{sec:P2}.
For inhomogeneous cases, in Sec.~\ref{sec:inhomo},
the stability of the normal ground state
is examined without the restriction of the dipole approximation.
The summary is shown in Sec.~\ref{sec:summary}.
The detailed derivation of the Dicke-like Hamiltonians is performed
in App.~\ref{sec:Dicke-like},
and transverse-longitudinal mixing in homogeneous systems
is discussed in App.~\ref{sec:LT-mixing}.

\section{Standard Hamiltonians in Coulomb and electric-dipole gauges}
\label{sec:standard}
For the system of charged particles in the presence of the radiation field,
the minimal-coupling Hamiltonian in the Coulomb gauge is expressed as \cite{cohen-tannoudji89}
\begin{align} \label{eq:oHC} 
\oHC
& = \sum_{\alpha}\frac{1}{2m_{\alpha}}\left[ \ovp_{\alpha} - q_{\alpha}\ovA(\ovr_{\alpha}) \right]^2
  + \oVC(\{\ovr_{\alpha}\})
\nonumber \\ & \quad
  + \frac{1}{2\diez} \intD\dd\vr\ \ovPL(\vr)^2 + \oHrad,
\end{align}
where the last term represents the energy of the radiation field:
\begin{equation} \label{eq:oHrad} 
\oHrad
= \intD\dd\vr\left\{ \frac{\ovPi(\vr)^2}{\diez} + \frac{[\rot\ovA(\vr)]^2}{\muz} \right\}.
\end{equation}
Here, $\diez$ and $\muz$ are the vacuum permittivity and vacuum permeability, respectively.
We in general suppose a finite space, and the subscript $\mathcal{D}$
means the integral in that space.
$\ovA(\vr)$ is the the vector potential,
and $\ovPi(\vr)$ is its conjugate momentum satisfying
$[ \ovA(\vr), \ovPi(\vr') ] = \ii\hbar\mdeltaT(\vr-\vr')$.
The dyadic transverse delta function $\mdeltaT(\vr)$ is defined
as \cite{cohen-tannoudji89}
\begin{equation} \label{eq:deltaT} 
\mdeltaT(\vr)
= \int\frac{\dd\vk}{(2\pi)^3}\left( \munit
  - \frac{\vk\vk}{k^2}
  \right) \ee^{\ii \vk\cdot\vr}.
\end{equation}
In the Coulomb gauge, the vector potential is transverse $\div\ovA(\vr) = 0$,
and the conjugate momentum corresponds to the
transverse electric field $\ovET(\vr) = - \ovPi(\vr)/\diez$.
The first term in Eq.~\eqref{eq:oHC} represents the kinetic energy of charged particles.
$\ovr_{\alpha}$ and $\ovp_{\alpha}$ are, respectively,
the position and momentum
of particle $\alpha$ with charge $q_{\alpha}$ and mass $m_{\alpha}$,
and they satisfy
$\left[ \ovr_{\alpha}, \ovp_{\alpha'} \right]
= \ii\hbar \delta_{\alpha,\alpha'} \munit$.
Expanding the kinetic term,
we get the light-matter interaction term
[$\ovp_{\alpha}\cdot\ovA(\ovr_{\alpha})$]
and also the $A^2$ term,
i.e., the square of the vector potential.
The second term $\oVC(\{\ovr_{\alpha}\})$ in Eq.~\eqref{eq:oHC}
represents the one-body potential
(e.g., external fields, core potential, etc.)
and many-body interaction of the particles
excluding the interaction between the longitudinal polarizations $\ovPL(\vr)$,
the third term in Eq.~\eqref{eq:oHC},
which corresponds to the (longitudinal) dipole-dipole interaction
discussed in Ref.~\cite{Keeling2007JPCM}.
In the case of infinite space,
we define the atomic polarization as \cite{cohen-tannoudji89}
\begin{equation}
\ovP(\vr) = \sum_{\alpha} \int_0^1\dd u\ q_{\alpha}\ovr_{\alpha}\delta(\vr-u\ovr_{\alpha}),
\end{equation}
and its transverse and longitudinal components are expressed as
\begin{subequations} \label{eq:ovPTL} 
\begin{align}
\ovPT(\vr) & = \int\dd\vr'\ \mdeltaT(\vr-\vr')\cdot\ovP(\vr'), \\
\ovPL(\vr) & = \int\dd\vr'\ \mdeltaL(\vr-\vr')\cdot\ovP(\vr'),
\label{eq:ovPL} 
\end{align}
\end{subequations}
where the longitudinal delta function is defined as
\cite{Note1}.
\begin{equation}
\mdeltaL(\vr)
= \delta(\vr)\munit - \mdeltaT(\vr)
= \int\frac{\dd\vk}{(2\pi)^3} \frac{\vk\vk}{k^2} \ee^{\ii \vk\cdot\vr}.
\end{equation}
Since $\ovPL(\vr)$ is defined only by the position operators $\{\ovr_{\alpha}\}$,
the Hamiltonian \eqref{eq:oHC} is included in those discussed
in the no-go theorem \cite{Bialynicki-Birula1979PRA,Rzazewski1991PRA},
and does not show the SPT under the dipole approximation in principle.

On the other hand, we can rewrite the Hamiltonian \eqref{eq:oHC}
by performing the Power-Zienau-Woolley (PZW) transformation,
which is shown in Ref.~\cite{cohen-tannoudji89} for infinite space
and is generalized to the finite case in Ref.~\cite{Vukics2014PRL}.
According to them, by neglecting the magnetic coupling between light and matter,
we get the multipolar-coupling Hamiltonian
(PZW gauge) as
\begin{align} \label{eq:oHP} 
\oHP
& = \sum_{\alpha}\frac{\ovp_{\alpha}{}^2}{2m_{\alpha}} + \oVC(\{\ovr_{\alpha}\})
+ \frac{1}{2\diez}\intD\dd\vr\ \ovP(\vr)^2
\nonumber \\ & \quad
- \frac{1}{\diez}\intD\dd\vr\ \ovPT(\vr) \cdot \ovDT(\vr)
+ \oHrad.
\end{align}
The energy $\oHrad$ of the radiation field is still expressed as
Eq.~\eqref{eq:oHrad}, while the conjugate momentum of the vector potential $\ovA(\vr)$
is the transverse electric displacement field
$\ovDT(\vr) = - \ovPi(\vr)$ in this gauge.
The kinetic energy of the particles is simply expressed as the first term,
and the light-matter interaction is described as the fourth term,
the product of the transverse atomic polarization $\ovPT(\vr)$
and the electric displacement $\ovDT(\vr)$.
Instead of the disappearance of the $A^2$ term,
we get the interaction between the transverse polarizations
$\intD\dd\vr\ \ovPT(\vr)^2/2\diez$,
which is called the $P^2$ term
and also prevents the SPT \cite{Emeljanov1976PLA,Yamanoi1978P2}.
Together with the interaction between longitudinal polarizations
in Eq.~\eqref{eq:oHC},
which survives even after the PZW transformation,
the interaction between the full polarizations
appears as the third term in Eq.~\eqref{eq:oHP}.
Due to the gauge-invariance,
the Hamiltonian \eqref{eq:oHP} does not also show the SPT
under the dipole approximation \cite{cohen-tannoudji89},
since it is denied in the Coulomb gauge by the no-go theorem
\cite{Bialynicki-Birula1979PRA,Rzazewski1991PRA}.

Let us derive the Dicke-like Hamiltonians from Eqs.~\eqref{eq:oHC} and \eqref{eq:oHP}.
For simplicity, we first consider a spatially infinite system,
and the vector potential and its conjugate momentum are expressed as
\cite{cohen-tannoudji89}
\begin{subequations}
\begin{align}
\ovA(\vr)
& = \sum_{\vk,\eta=1,2} \vunit_{\vk,\eta} \sqrt{\frac{\hbar}{2\diez c|\vk|\vol}}
    \left( \oa_{\vk,\eta} + \oad_{-\vk,\eta} \right) \ee^{\ii\vk\cdot\vr}, \\
\ovPi(\vr)
& = - \sum_{\vk,\eta=1,2} \vunit_{\vk,\eta} \ii\sqrt{\frac{\hbar\diez c|\vk|}{2\vol}}
    \left( \oa_{\vk,\eta} - \oad_{-\vk,\eta} \right) \ee^{\ii\vk\cdot\vr}.
\end{align}
\end{subequations}
Here, $c$ is the speed of light and $\vol \rightarrow \infty$ is the volume of the space.
$\vunit_{\vk,\eta}$ is the unit vector in the two directions ($\eta = 1,2$)
perpendicular to wavevector $\vk$.
$\oa_{\vk,\eta}$ is the annihilation operator of a photon
with wavevector $\vk$ and polarization $\eta$.
We also simply suppose that all the atoms are identical,
and they spread homogeneously in the whole space.
Further, each atom has an isotropic transition from its ground state $\ket{g}$
to an excited state $\ket{e}$ with transition frequency $\wa$.

Here, we assume that the system is near the normal ground state,
where the radiation and polarization fields have no amplitude.
In this situation,
since the system is infinite, homogeneous, and isotropic (IHI),
we do not lose the generality by focusing on one direction of wavevector $\vk$.
The two transverse fields and one longitudinal field
are defined in that direction.
For discussing small deviations from the normal ground state,
the transverse and longitudinal fields are all decoupled
(see App.~\ref{sec:LT-mixing}).
Then, we can separately discuss one of the transverse fields ($\eta = 1$).
Further, we focus on only the two states $\ket{g}$ and $\ket{e}$ in each atom
(two-level approximation).
The atomic part of the Hamiltonian is then simplified as
\begin{equation}
\oHatom = \sum_{\alpha}\frac{ \ovp_{\alpha}{}^2}{2m_{\alpha}} + \oVC(\{\ovr_{\alpha}\})
\rightarrow \sum_{\lambda=1}^N \frac{\hbar\wa}{2}\osigma_{\lambda}^z,
\end{equation}
where $\osigma_{\lambda}^z$ is the Pauli matrix
representing the population of atom $\lambda$,
and $N$ is the number of atoms.
Under the dipole approximation (long-wavelength approximation;
$\ee^{\ii\vk\cdot\ovr_{\alpha}} \simeq \ee^{\ii\vk\cdot\vR_{\lambda}}$
for the short distance $|\ovr_{\alpha}-\vR_{\lambda}|$ in each atom
placing at $\vR_{\lambda}$),
the minimal-coupling Hamiltonian \eqref{eq:oHC} and multipolar-coupling
one \eqref{eq:oHP} are rewritten, respectively, as
(see App.~\ref{sec:Dicke-like})
\begin{subequations} \label{eq:oH1D} 
\begin{align} \label{eq:oHCO} 
\oHCO
& = \sum_{k} \hbar c|k| \oad_{k}\oa_{k}
+ \sum_{\lambda=1}^N \frac{\hbar\wa}{2} \osigma_{\lambda}^z
\nonumber \\ & \quad
+ \sum_k \sum_{\lambda=1}^N \frac{\hbar\wa\rabi_{k}}{\sqrt{N}}
  \osigma_{\lambda}^y \left( \oa_{k} + \oad_{-k} \right) \ee^{\ii kR_{\lambda}}
\nonumber \\ & \quad
+ \sum_{k} \hbar\wa\rabi_{k}{}^2 \left( \oa_{k} + \oad_{-k} \right)
   \left( \oa_{-k} + \oad_{k} \right),
\end{align}
\begin{align}
\oHPO
& = \sum_{k} \hbar c|k| \oad_{k}\oa_{k}
+ \sum_{\lambda=1}^N\frac{\hbar\wa}{2}\osigma_{\lambda}^z
\nonumber \\ & \quad
- \sum_k \sum_{\lambda=1}^N \frac{\ii\hbar c|k|\rabi_{k}}{\sqrt{N}}
  \osigma_{\lambda}^x \left( \oa_{k} - \oad_{-k} \right)
  \ee^{\ii kR_{\lambda}}
\nonumber \\ & \quad
+ \sum_k \sum_{\lambda=1}^N \sum_{\lambda'=1}^N
  \frac{\hbar c|k|\rabi_{k}{}^2}{N}
  \osigma_{\lambda}^x \osigma_{\lambda'}^x
  \ee^{\ii k(R_{\lambda}-R_{\lambda'})}.
\label{eq:oHPO} 
\end{align}
\end{subequations}
Here, the Pauli matrices are defined as
$\osigma_{\lambda}^x = \osigma_{\lambda} + \osigmad_{\lambda}$
and $\osigma_{\lambda}^y = \ii(\osigma_{\lambda} - \osigmad_{\lambda})$
for the lowering operator $\osigma_{\lambda}$ of atom $\lambda$.
The non-dimensional light-matter coupling strength is obtained as
\begin{equation}
\rabi_{k}
= \sqrt{\frac{N|\dm|^2}{2\hbar\diez c|k|\vol}}.
\end{equation}
The dipole moment $\dm$ is defined as
\begin{equation}
\dm = \sum_{\alpha\ \text{in an atom}}
\braket{e|q_{\alpha} \ovr_{\alpha}|g} \cdot \vunit_{\vk,1},
\end{equation}
where the summation is over the charged particles $\{\alpha\}$ localized in an atom.
The PZW gauge is reduced to the electric-dipole gauge
under the dipole (long-wavelength) approximation \cite{cohen-tannoudji89,Keeling2007JPCM,Vukics2014PRL}.
The last terms in Eqs.~\eqref{eq:oHCO} and \eqref{eq:oHPO}
are derived from the $A^2$ term in Eq.~\eqref{eq:oHC}
and (transverse) $P^2$ term in Eq.~\eqref{eq:oHP}, respectively.
When we perform the single-mode approximation by supposing
a cavity of radiation field without mixing transverse and longitudinal fields,
the two Hamiltonians \eqref{eq:oHCO} and \eqref{eq:oHPO}
are reduced to the Dicke Hamiltonians \eqref{eq:oHDicke}
with the $A^2$ term \eqref{eq:oHA2} and with the $P^2$ term \eqref{eq:oHP2},
respectively,
while the counter-rotating terms $\oa\osigma_{\lambda}$ and $\oad\osigmad_{\lambda}$
also appear.
Due to the presence of the $A^2$ or $P^2$ term,
the SPT is prevented in either gauge
regardless of whether the radiation field consists of single mode
\cite{Rzazewski1975PRL,Knight1978PRA,Bialynicki-Birula1979PRA}
or continuous ones \cite{Rzazewski1976PRA,Yamanoi1976PLA,Emeljanov1976PLA,Yamanoi1978P2,Yamanoi1979JPA}.
Since this result is consistent with the assumption of the normal ground state,
it is then found to be stable in the IHI systems.
Whereas this logic does not deny the possibility of another stable state
with non-zero field amplitudes,
it is known that the SPT accompanies with the instability of the normal ground state
in the IHI systems \cite{Yamanoi1976PLA,Emeljanov1976PLA,Yamanoi1978P2,Yamanoi1979JPA}
even with keeping all the transverse and longitudinal fields \cite{Yamanoi1978P2}.
Further, this result is consistent also with the no-go theorem
\cite{Bialynicki-Birula1979PRA,Rzazewski1991PRA}.

\section{Elimination of $P^2$ term in homogeneous and isotropic systems}
\label{sec:P2}
Although the SPT is prevented in the above scenario,
we can eliminate the $P^2$ term in the PZW gauge
by the treatment discussed in $\mathrm{C_{IV}}$ of Ref.~\cite{cohen-tannoudji89}
and as pointed out in Refs.~\cite{Keeling2007JPCM,Vukics2014PRL}.
Obeying this scenario, the SPT seems to occur at first glance.
This treatment can be applied when the polarization in each atom
is well localized
and they have no overlap with neighboring atoms,
i.e., when we express the total polarization density $\ovP(\vr)$ by
individual atomic polarization $\ovP_{\lambda}(\vr)$ as
\begin{equation}
\ovP(\vr) = \sum_{\lambda} \ovP_{\lambda}(\vr),
\end{equation}
\begin{equation}
\ovP_{\lambda}(\vr) = \sum_{\alpha\ \text{in atom $\lambda$}}
\int_0^1\dd u\ q_{\alpha}\ovr_{\alpha}\delta(\vr-u\ovr_{\alpha}),
\end{equation}
the overlap between different atomic polarizations are supposed to be zero as
\begin{equation}
\int\dd\vr\ \ovP_{\lambda}(\vr) \cdot \ovP_{\lambda'}(\vr)
= 0 \quad \text{for $\lambda \neq \lambda'$}.
\end{equation}
Under this assumption,
we can eliminate the inter-atomic interactions
in the full $P^2$ term in Eq.~\eqref{eq:oHP} as
\begin{equation}
\frac{1}{2\diez}\intD\dd\vr\ \ovP(\vr)^2
= \sum_{\lambda=1}^N \frac{1}{2\diez}\intD\dd\vr\ \ovP_{\lambda}(\vr)^2.
\end{equation}
The remaining intra-atomic interaction (self energy) in each atom
can be renormalized to $\oHatom$,
and then the full $P^2$ term in Eq.~\eqref{eq:oHP} can be eliminated.
In this way, the $P^2$ term in Eq.~\eqref{eq:oHPO} seems to disappear,
and the SPT seems to be enabled by this treatment at first glance.
Even in the Coulomb gauge, the longitudinal $P^2$ term in Eq.~\eqref{eq:oHC}
can be rewritten as
\begin{equation} \label{eq:PL2=P2-PT2} 
\frac{1}{2\diez}\intD\dd\vr\ \ovPL(\vr)^2
= \frac{1}{2\diez}\intD\dd\vr \left[
  \sum_{\lambda=1}^N \ovP_{\lambda}(\vr)^2 - \ovPT(\vr)^2
  \right].
\end{equation}
Due to the presence of the transverse $P^2$ term (the second term on the right hand side)
and the $A^2$ term, the same kind of phase transition occurs
as pointed out in Ref.~\cite{Keeling2007JPCM}.
The SPT in a similar Hamiltonian
with the transverse dipole-dipole interaction
was examined also in Ref.~\cite{Knight1978PRA}.

In contrast, as we derived Eqs.~\eqref{eq:oH1D},
after we extract one of the transverse fields
(truncating the other fields),
which can be justified in the IHI system (no transverse-longitudinal mixing),
the absence of the inter-atomic overlap can modify the light-matter coupling strength $\rabi_k$,
but it never eliminates the $A^2$ or transverse $P^2$ term
in Eqs.~\eqref{eq:oH1D}, which still prevent the SPT.
In other words, we can no longer eliminate the $P^2$ term in the PZW gauge
after extracting the transverse field.

In this way, concerning the possibility of the SPT,
we get inconsistent results depending on whether
the elimination of (inter-atomic) full $P^2$ term
is performed before the extraction of a transverse field.
This elimination is the trick to avoid the no-go theorem
\cite{Bialynicki-Birula1979PRA,Rzazewski1991PRA},
and the problem is whether it can be justified
for deriving the Dicke-like Hamiltonians \eqref{eq:oH1D} or not.
At least for the IHI system,
since the assumption of no inter-atomic overlap does not destroy
the homogeneity or the isotropy of the system,
the transverse and longitudinal fields are still decoupled
near the normal ground state.
Then, the extraction of the transverse field itself can be justified
even under the no-overlap assumption.

In contrast, the elimination of the full $P^2$ term
can be justified only in the presence of all
the transverse and longitudinal fields for all $\vk$,
because this treatment gives the energy balance
between the (inter-atomic) transverse $P^2$ term
and the longitudinal one as seen in Eq.~\eqref{eq:PL2=P2-PT2}.
Thanks to this energy-balance relation,
this treatment effectively mixes
the transverse and longitudinal fields
(and modifies the Hamiltonian from that discussed in the no-go theorem
\cite{Bialynicki-Birula1979PRA,Rzazewski1991PRA}),
although they are decoupled in the IHI systems
even under the no-overlap assumption.
If we truncate a part of the fields that originally compose the system,
the energy balance can no longer be considered properly.

From the mathematical viewpoint,
the elimination by the no-overlap assumption is justified only for the field
defined locally in space such as the full polarization $\ovP(\vr)$.
Although the full $P^2$ term is reduced to the intra-atomic interaction
as $\ovP_{\lambda}(\vr)^2$ in Eq.~\eqref{eq:PL2=P2-PT2},
the $\ovPT(\vr)^2$ and $\ovPL(\vr)^2$ terms keep their forms
even under the no-overlap assumption.
As is also mentioned in Ref.~\cite{Keeling2007JPCM},
this is because the transverse and longitudinal fields $\ovPT(\vr)$
and $\ovPL(\vr)$ are defined
non-locally in space by the transverse and longitudinal delta functions
as Eqs.~\eqref{eq:ovPTL}.
If we finally want to extract one of the transverse fields
for deriving the Dicke-like Hamiltonians \eqref{eq:oH1D},
we should not eliminate the full $P^2$ term,
because this treatment is fragile against the non-locality,
which emerges by the extraction.
Then, after the elimination of the full $P^2$ term,
we must keep all the transverse and longitudinal fields
for discussing the SPT as in Ref.~\cite{Yamanoi1978P2}.

In this way, at least in the IHI systems starting from the minimal-coupling Hamiltonian
\eqref{eq:oHC} or the multipolar-coupling one \eqref{eq:oHP},
the SPT is forbidden as proved by the no-go theorem
\cite{Bialynicki-Birula1979PRA,Rzazewski1991PRA},
and the elimination of the $P^2$ term must be performed with an attention
to its fragility against the non-locality.

\section{Stability of normal ground state in inhomogeneous systems}
\label{sec:inhomo}
As discussed in the previous section,
the presence of the longitudinal dipole-dipole interaction and
the absence of the inter-atomic overlap does not enable the SPT
at least when the transverse and longitudinal fields are decoupled
in the IHI systems.
However, there still remains the possibility of the SPT
in inhomogeneous (finite) systems
such as cavity structures or anisotropic systems,
in which the transverse and longitudinal fields are generally mixed.
In the rest of this paper, we show that the inhomogeneity or system-finiteness does not destabilize
the normal ground state of the generally anisotropic polarizable materials.

In the case of the IHI systems,
Refs.~\cite{Yamanoi1976PLA,Emeljanov1976PLA,Yamanoi1978P2,Yamanoi1979JPA}
calculated the optical susceptibility for the normal ground state,
and examined the stability against small deviations of the radiation and polarization amplitudes
by calculating the poles of the system.
The authors found poles in the upper half complex plane
at the conditions that the superradiant state appears,
i.e., the normal ground state becomes unstable at the same conditions
as the SPT.

Extending their discussions, we examine the stability of the normal ground state in inhomogeneous and anisotropic systems.
The following discussion is not restricted by the dipole approximation
used in the no-go theorem \cite{Bialynicki-Birula1979PRA,Rzazewski1991PRA}.
First of all, we suppose that the system does not show the SPT at least for the homogeneous and infinite case,
which should be checked by microscopic calculations.
For the normal ground state of the homogeneous and infinite system,
we can calculate the optical susceptibility $\mchi(\omega)$ 
and the dielectric function $\mdie(\omega) = \munit + \mchi(\omega)$,
which are in general dyadic for anisotropic systems \cite{Note3}.
They connects
the small deviations of the polarization and electric field
from the normal ground state [$\vP(\vr) = \vE(\vr) = 0$]
in the classical electrodynamics in the frequency domain as
\begin{equation}
\vP(\vr,\omega) = \diez\mchi(\omega)\cdot\vE(\vr,\omega).
\end{equation}
Here, we suppose that such polarizable materials spread inhomogeneously
in finite or infinite space
without modifying the optical susceptibility $\mchi(\omega)$
determined microscopically without any microscopic surface effects.
Then, the polarizable materials can be characterized
simply by the position-dependent susceptibility $\mchi(\vr,\omega)$
or the dielectric function $\mdie(\vr,\omega) = \munit + \mchi(\vr,\omega)$.
The finite systems such as in a cavity can also be considered
by supposing perfect mirrors through $\mdie(\vr,\omega)$.
We try to examine the stability of the normal ground state
in such inhomogeneous polarizable systems.

The quantum electrodynamics of dispersive, absorptive, inhomogeneous,
and anisotropic materials can be discussed
in the framework of Ref.~\cite{knoll01}
including the longitudinal $P^2$ term.
The atomic excitations discussed in the Dicke Hamiltonian
can be simplified to bosonic oscillators
supposed in Ref.~\cite{knoll01}
when we consider the small deviations from the normal ground state
\cite{Yamanoi1976PLA,Emeljanov1976PLA,Yamanoi1978P2,Yamanoi1979JPA}.
As a result in this framework and also discussed in Ref.~\cite{abrikosov75ch6},
the retarded Green function of the electric field
in the inhomogeneous and anisotropic system is obtained as
\begin{equation} \label{eq:<EE>=G} 
\int_0^{\infty}\dd t\ \ee^{\ii\omega t}
\braket{[ \ovE(\vr,t), \ovE(\vr',0)]}
= - \ii\hbar\muz\omega^2\mG(\vr,\vr',\omega),
\end{equation}
where the dyadic function $\mG(\vr,\vr',\omega)$ satisfies
the following wave equation characterized by $\mdie(\vr,\omega)$:
\begin{equation}
\rot\rot\mG(\vr,\vr',\omega)
- \frac{\omega^2}{c^2}\mdie(\vr,\omega)\cdot\mG(\vr,\vr',\omega)
= \delta(\vr-\vr')\munit.
\end{equation}
The expectation value on the left hand side in Eq.~\eqref{eq:<EE>=G} is taken for the normal ground state
of the system as is used for evaluating $\mdie(\vr,\omega)$.
Then, we can examine the stability of the normal ground state
by the poles of the Green function $\mG(\vr,\vr',\omega)$.
As discussed in Ref.~\cite{knoll01}, even in the inhomogeneous system,
the Green function has no pole in the upper half plane
if the dielectric function $\mdie(\vr,\omega)$ satisfies
the same condition.
This is simply due to the causality.
Therefore, if the normal ground state is stable (has no SPT) in the homogeneous case,
the inhomogeneity itself does not destabilize the normal ground state,
although it connects macroscopically the transverse and longitudinal fields,
which were missing in the former discussion of this paper.
Whereas this logic does not also deny the possibility of another stable state,
the SPT accompanies the instability of the normal ground state
at least for the homogeneous case.

\section{Summary}
\label{sec:summary}
We examined the stability of the normal ground state (having no field amplitude)
of polarizable materials
described by Eq.~\eqref{eq:oHC} or \eqref{eq:oHP}
under focusing especially on the longitudinal and transverse dipole-dipole interactions
and the validity of eliminating the full $P^2$ term,
which were pointed out in Refs.~\cite{Keeling2007JPCM,Vukics2014PRL}.
We conclude that
the longitudinal dipole-dipole interaction
in the absence of the inter-atomic overlap does not enable the SPT
for the infinite, homogeneous, and isotropic systems.
This is because the elimination of the full $P^2$ term in the electric-dipole gauge
can be justified only when all the transverse and longitudinal fields remain.
If we want to eliminate the full $P^2$ term,
the possibility of the SPT should be discussed
with keeping all those fields.
Once the SPT is found to be forbidden in homogeneous systems
by microscopic calculations,
the stability of the normal ground state
in inhomogeneous cases can be examined by the poles of the system.
It is found without the restriction of the dipole approximation
that the macroscopic inhomogeneity itself does not destabilize the normal ground state.

Then, when we investigate the systems with the ultrastrong light-matter coupling,
if they are proved not to show the SPT in equilibrium situations,
we should consider the $A^2$ or $P^2$ term in the Hamiltonian
or it should be renormalized to the cavity frequency $\wc$
or atomic excitation one $\wa$, respectively.
The shifted cavity or atomic frequency also prevents the SPT.
In the non-equilibrium situations, we can obtain the analogues of the SPT
even in the presence of the $A^2$ or $P^2$ term.
In the equilibrium situations,
the possibility of the SPT should be discussed
in infinite and homogeneous systems
from microscopic calculations,
or microscopic surface effects on the atomic transitions
must be introduced explicitly in inhomogeneous systems.
Such microscopic effects must be the ones beyond the model in the no-go theorem \cite{Bialynicki-Birula1979PRA,Rzazewski1991PRA},
e.g., the magnetic-dipole interaction
involving electron spins \cite{Knight1978PRA},
which should be examined probably based on the Dirac equation.

\begin{acknowledgments}
M.B.~thanks J.~Keeling, A.~Vukics, T.~Chartier, T.~Shirai, K.~Asano,
and T.~Yuge for fruitful discussions.
This work was funded by ImPACT Program of Council for Science, Technology and
Innovation (Cabinet Office, Government of Japan)
and by KAKENHI (No.~26287087 and No.~24-632).
\end{acknowledgments}

\appendix
\section{Derivation of Dicke-like Hamiltonians}
\label{sec:Dicke-like}
From the original Hamiltonians \eqref{eq:oHC}
and \eqref{eq:oHP} in the Coulomb and PZW gauges,
here we derive the Dicke-like Hamiltonians under the two-level
and long-wavelength (dipole) approximations.
Whereas we finally reduce the atomic states into the two levels
(ground and excited states),
we suppose that the excited states actually have the three-fold degeneracy
in the three-dimensional and isotropic systems.
This is because the electric-dipole transition is allowed
only between two states with different parities
(e.g., between $s$- and $p$-orbitals).
When we suppose isotropic atoms and the ground state $\ket{g}$ is not degenerated,
the excited states $\ket{\xi}$ in each atom have a degree of freedom of the orientation
$\xi = \{1,2,3\}$ orthogonal to each other.
Here, we introduce extended Pauli matrices
$\osigma_{\lambda,\xi}^{x,y}$ and $\osigma_{\lambda}^z$ for each atom
(identified by index $\lambda = 1 \ldots N$).
Since the three excited states in each atom share the same ground state,
these states are expressed by the following $4\times4$ matrices for each atom as
\begin{subequations}
\begin{align}
\osigma_{\xi}
& = \begin{pmatrix}
0 & 0 & 0 & 0 \\
0 & 0 & 0 & 0 \\
0 & 0 & 0 & 0 \\
\delta_{\xi,1} & \delta_{\xi,2} & \delta_{\xi,3} & 0
\end{pmatrix}, \\
\osigma^x_{\xi} & = \osigma_{\xi} + \osigmad_{\xi}, \\
\osigma^y_{\xi} & = \ii( \osigma_{\xi} - \osigmad_{\xi} ), \\
\osigma^z & = \begin{pmatrix}
1 & 0 & 0 & 0 \\
0 & 1 & 0 & 0 \\
0 & 0 & 1 & 0 \\
0 & 0 & 0 & -1
\end{pmatrix},
\end{align}
\end{subequations}
where the basis of the matrices is $\{\ket{1}, \ket{2}, \ket{3}, \ket{g}\}$.

Using these operators,
the atomic polarization is expressed
under the two-level (but three-fold degeneracy for excited states)
and long-wavelength (dipole) approximations as
\begin{equation}
\int\dd\vr\ \ovP(\vr) \ee^{\ii\vk\cdot\vr}
= \sum_{\xi=1,2,3} \vunit_{\xi} \dm \sum_{\lambda=1}^N
  \osigma_{\lambda,\xi}^x \ee^{\ii\vk\cdot\vR_{\lambda}},
\end{equation}
where $\vunit_{\xi}$ is the unit vector in the $\xi$ orientation.
In the same manner, we can also rewrite $\ovp_{\alpha}$ in $\oHC$,
and then the Hamiltonians \eqref{eq:oHC}
and \eqref{eq:oHP} are rewritten as
\begin{widetext}
\begin{subequations} \label{eq:oH_a_Pauli} 
\begin{align} \label{eq:oHC_a_Pauli} 
\oHC
& = \sum_{\vk}\sum_{\eta=1,2} \hbar c|\vk| \oad_{\vk,\eta}\oa_{\vk,\eta}
  + \sum_{\lambda=1}^N \frac{\hbar\wa}{2}\osigma_{\lambda}^z
\nonumber \\ & \quad
+ \sum_{\vk}\sum_{\eta=1,2}\sum_{\lambda=1}^N\sum_{\xi=1,2,3}
  \vunit_{\xi}\cdot\vunit_{\vk,\eta}
  \frac{\hbar\wa\rabi_k}{\sqrt{N}}
  \osigma_{\lambda,\xi}^y
  \left( \oa_{\vk,\eta} + \oad_{-\vk,\eta} \right)
  \ee^{\ii\vk\cdot\vR_{\lambda}}
\nonumber \\ & \quad
+ \sum_{\vk}\sum_{\eta=1,2} \hbar\wa\rabi_k{}^2 \left( \oa_{\vk,\eta} + \oad_{-\vk,\eta} \right)
   \left( \oa_{-\vk,\eta} + \oad_{\vk,\eta} \right)
\nonumber \\ & \quad
+ \sum_{\vk}\sum_{\lambda=1}^N\sum_{\xi=1,2,3}
  \sum_{\lambda'=1}^N\sum_{\xi'=1,2,3}
  \vunit_{\xi}\cdot\vunit_{\vk,3}
  \vunit_{\xi'}\cdot\vunit_{\vk,3}
  \frac{\hbar c|k|\rabi_k{}^2}{N}
  \osigma_{\lambda,\xi}^x \osigma_{\lambda',\xi'}^x
  \ee^{\ii\vk\cdot(\vR_{\lambda}-\vR_{\lambda'})},
\end{align}
\begin{align}
\oHP
& = \sum_{\vk}\sum_{\eta=1,2} \hbar c|\vk| \oad_{\vk,\eta}\oa_{\vk,\eta}
  + \sum_{\lambda=1}^N\frac{\hbar\wa}{2}\osigma_{\lambda}^z
\nonumber \\ & \quad
- \sum_{\vk}\sum_{\eta=1,2}\sum_{\lambda=1}^N\sum_{\xi=1,2,3}
  \vunit_{\xi}\cdot\vunit_{\vk,\eta}
  \frac{\ii\hbar c|k|\rabi_k}{\sqrt{N}}
  \osigma_{\lambda,\xi}^x
  \left( \oa_{\vk,\eta} - \oad_{-\vk,\eta} \right)
  \ee^{\ii\vk\cdot\vR_{\lambda}}
\nonumber \\ & \quad
+ \sum_{\vk}\sum_{\eta=1,2,3}\sum_{\lambda=1}^N\sum_{\xi=1,2,3}
  \sum_{\lambda'=1}^N\sum_{\xi'=1,2,3}
  \vunit_{\xi}\cdot\vunit_{\vk,\eta}
  \vunit_{\xi'}\cdot\vunit_{\vk,\eta}
  \frac{\hbar c|k|\rabi_k{}^2}{N}
  \osigma_{\lambda,\xi}^x \osigma_{\lambda',\xi'}^x
  \ee^{\ii\vk\cdot(\vR_{\lambda}-\vR_{\lambda'})},
 \label{eq:oHP_a_Pauli} 
\end{align}
\end{subequations}
\end{widetext}
where $\vunit_{\vk,3} = \vk/|k|$ is the unit vector parallel to $\vk$.
For deriving the former equation in the Coulomb gauge, we used the following relation 
\begin{equation}
\braket{e|\ovp_{\alpha}|g}
= \frac{m_{\alpha}}{\ii\hbar} \braket{e|[\ovr_{\alpha}, \oHatom]|g}
= \frac{\ii m_{\alpha} \wa}{q_{\alpha}}\braket{e|q_{\alpha} \ovr_{\alpha}|g},
\end{equation}
and the sum rule of the oscillator strengths
simplified under the two-level approximation
\begin{equation}
\frac{2\wa|\dm|^2}{\hbar}
\left(\sum_{\alpha\ \text{in an atom}} \frac{m_{\alpha}}{q_{\alpha}{}^2}\right)^{-1}
= 1,
\end{equation}
which is derived from 
\begin{equation}
\frac{m_{\alpha}}{\hbar^2}\left[ \ovr_{\alpha}, \left[ \oHatom, \ovr_{\alpha'} \right] \right]
= \delta_{\alpha,\alpha'}\munit.
\end{equation}

The last term in Eq.~\eqref{eq:oHP_a_Pauli} is the full $P^2$ term,
and it is simply rewritten as
\begin{equation}
\sum_{\lambda=1}^N\sum_{\xi=1,2,3}
  \frac{N|\dm|^2}{2\diez\vol}
  \osigma_{\lambda,\xi}^x \osigma_{\lambda,\xi}^x.
\end{equation}
Then, this term can be reduced to the intra-atomic interaction
as discussed in Sec.~\ref{sec:P2}.
However, the expression in Eq.~\eqref{eq:oHP_a_Pauli}
is more useful in order to derive the Dicke-like Hamiltonians.
Thanks to the homogeneity and isotropy,
we can focus on one direction of $\vk$
without losing the generality,
and determine two directions $\eta=\{1,2\}$ perpendicular to $\vk$.
Along the two transverse directions $\eta$,
we can also define $\xi=\{1,2\}$ perpendicular to $\vk$
and $\xi=3$ parallel to $\vk$.
Focusing on one of the transverse field in the fixed direction of $\vk$,
Eqs.~\eqref{eq:oH_a_Pauli} are reduced to the Dicke-like Hamiltonians $\oHCO$ and $\oHPO$
shown in Eqs.~\eqref{eq:oH1D}.
The next problem is how we can justify this extraction of the subsystem.
In the next section, we discuss the mixing of transverse and longitudinal fields
in the ensemble of atoms.

\section{Mixing of transverse and longitudinal fields}
\label{sec:LT-mixing}
In fact, in the Hamiltonians \eqref{eq:oH_a_Pauli},
all the transverse and longitudinal fields are mixed in general,
i.e., the Dicke-like Hamiltonian $\oHCO$ ($\oHPO$), Eqs.~\eqref{eq:oH1D},
is not commutable with the rest of the Hamiltonian in either gauge.
However, for discussing small deviations from the normal ground state
(no field amplitude), we can approximately bosonize the two-level atomic system.
First, we introduce the following operator that annihilates
an excitation with wavevector $\vk$ and polarization $\xi$:
\begin{equation}
\ob_{\vk,\xi} = \frac{1}{\sqrt{N}} \sum_{\lambda=1}^N \osigma_{\lambda,\xi} \ee^{-\ii\vk\cdot\vR_{\lambda}}.
\end{equation}
The Hamiltonians \eqref{eq:oH_a_Pauli} are rewritten without any approximation as
\begin{subequations}
\begin{align}
\oHC & = \sum_{\vk} \oHC^{\vk}, \\
\oHP & = \sum_{\vk} \oHP^{\vk},
\end{align}
\end{subequations}
where the subsystems characterized by wavevector $\vk$ is expressed as
\begin{widetext}
\begin{subequations}
\begin{align}
\oHC^{\vk}
& = \sum_{\eta=1,2} \hbar c|\vk| \oad_{\vk,\eta}\oa_{\vk,\eta}
  + \sum_{\xi=1,2,3} \hbar\wa\obd_{\vk,\xi}\ob_{\vk,\xi} + \const
\nonumber \\ & \quad
+ \sum_{\eta=1,2}\sum_{\xi=1,2,3}
  \vunit_{\xi}\cdot\vunit_{\vk,\eta}
  \ii\hbar\wa\rabi_k
  \left( \oa_{\vk,\eta} + \oad_{-\vk,\eta} \right)
  \left( \ob_{-\vk,\xi} - \obd_{\vk,\xi} \right)
\nonumber \\ & \quad
+ \sum_{\eta=1,2} \hbar\wa\rabi_k{}^2 \left( \oa_{\vk,\eta} + \oad_{-\vk,\eta} \right)
   \left( \oa_{-\vk,\eta} + \oad_{\vk,\eta} \right)
\nonumber \\ & \quad
+ \sum_{\xi=1,2,3}\sum_{\xi'=1,2,3}
  \vunit_{\xi}\cdot\vunit_{\vk,3}
  \vunit_{\xi'}\cdot\vunit_{\vk,3}
  \hbar c|k|\rabi_k{}^2
  \left( \ob_{\vk,\xi} + \obd_{-\vk,\xi} \right)
  \left( \ob_{-\vk,\xi'} + \obd_{\vk,\xi'} \right),
\end{align}
\begin{align}
\oHP^{\vk}
& = \sum_{\eta=1,2} \hbar c|\vk| \oad_{\vk,\eta}\oa_{\vk,\eta}
+ \sum_{\xi=1,2,3} \hbar\wa\obd_{\vk,\xi}\ob_{\vk,\xi} + \const
\nonumber \\ & \quad
- \sum_{\eta=1,2}\sum_{\xi=1,2,3}
  \vunit_{\xi}\cdot\vunit_{\vk,\eta}
  \ii\hbar c|k|\rabi_k
  \left( \oa_{\vk,\eta} - \oad_{-\vk,\eta} \right)
  \left( \ob_{-\vk,\xi} + \obd_{\vk,\xi} \right)
\nonumber \\ & \quad
+ \sum_{\eta=1,2,3}\sum_{\xi=1,2,3}\sum_{\xi'=1,2,3}
  \vunit_{\xi}\cdot\vunit_{\vk,\eta}
  \vunit_{\xi'}\cdot\vunit_{\vk,\eta}
  \hbar c|k|\rabi_k{}^2
  \left( \ob_{\vk,\xi} + \obd_{-\vk,\xi} \right)
  \left( \ob_{-\vk,\xi'} + \obd_{\vk,\xi'} \right).
\end{align}
\end{subequations}
\end{widetext}
Here, since the system is homogeneous, the commutator of $\ob_{\vk,\xi}$ is derived as
\begin{align}
& \left[ \ob_{\vk,\xi}, \obd_{\vk',\xi'} \right]
= \delta_{\vk,\vk'}\delta_{\xi,\xi'}
\nonumber \\ & \quad
- \frac{\delta_{\xi,\xi'}}{N} \sum_{\lambda=1}^N \left(
   2\osigmad_{\lambda,\xi}\osigma_{\lambda,\xi}
  + \sum_{\xi''\neq\xi}\osigmad_{\lambda,\xi''}\osigma_{\lambda,\xi''}
\right)\ee^{-\ii(\vk-\vk')\cdot\vR_{\lambda}}.
\end{align}
If the expectation number of excitations is quite smaller than the number of atoms $N$,
the first term is the major contribution, and the others are negligible.
Then, for discussing the small deviations from the normal ground state,
we can bosonize the excitation operators as
\begin{equation}
\left[ \ob_{\vk,\xi}, \obd_{\vk',\xi'} \right]
\simeq \delta_{\vk,\vk'}\delta_{\xi,\xi'}.
\end{equation}
Under this approximation, in either gauge,
the Hamiltonians $\oHC^{\vk}$ ($\oHP^{\vk}$)
are commutable with each other for different wavevector
(except for $-\vk$, but it is in the same direction).

We next decompose the subsystem of $\vk$
into three subsystems: two transverse fields
and one longitudinal field.
For the fixed $\vk$, we take the same orientations
for the radiation and polarization fields
($\vunit_{\vk,\eta}\cdot\vunit_{\xi} = \delta_{\eta,\xi}$
and $\vunit_3 = \vk/|k|$).
Then, the Hamiltonians are rewritten as
\begin{subequations}
\begin{align}
\oHC^{\vk} & = \sum_{\xi=1,2,3} \oHC^{\vk,\xi}, \\
\oHP^{\vk} & = \sum_{\xi=1,2,3} \oHP^{\vk,\xi},
\end{align}
\end{subequations}
where
\begin{subequations}
\begin{align}
\oHC^{\vk,\xi=1,2}
& = \hbar c|\vk| \oad_{\vk,\xi}\oa_{\vk,\xi}
  + \hbar\wa\obd_{\vk,\xi}\ob_{\vk,\xi} + \const
\nonumber \\ & \quad
+ \ii\hbar\wa\rabi_k
  \left( \oa_{\vk,\xi} + \oad_{-\vk,\xi} \right)
  \left( \ob_{-\vk,\xi} - \obd_{\vk,\xi} \right)
\nonumber \\ & \quad
+ \hbar\wa\rabi_k{}^2 \left( \oa_{\vk,\xi} + \oad_{-\vk,\xi} \right)
   \left( \oa_{-\vk,\xi} + \oad_{\vk,\xi} \right),
\label{eq:oHCT} 
\end{align}
\begin{align}
\oHP^{\vk,\xi=1,2}
& = \hbar c|\vk| \oad_{\vk,\xi}\oa_{\vk,\xi}
  + \hbar\wa\obd_{\vk,\xi}\ob_{\vk,\xi} + \const
\nonumber \\ & \quad
- \ii\hbar c|k|\rabi_k
  \left( \oa_{\vk,\xi} - \oad_{-\vk,\xi} \right)
  \left( \ob_{-\vk,\xi} + \obd_{\vk,\xi} \right)
\nonumber \\ & \quad
+ \hbar c|k|\rabi_k{}^2
  \left( \ob_{\vk,\xi} + \obd_{-\vk,\xi} \right)
  \left( \ob_{-\vk,\xi} + \obd_{\vk,\xi} \right),
\label{eq:oHPT} 
\end{align}
\begin{align}
\oHC^{\vk,3}
& = \oHP^{\vk,3} \nonumber \\
& = \hbar\wa\obd_{\vk,3}\ob_{\vk,3} + \const
\nonumber \\ & \quad
+ \hbar c|k|\rabi_k{}^2
  \left( \ob_{\vk,3} + \obd_{-\vk,3} \right)
  \left( \ob_{-\vk,3} + \obd_{\vk,3} \right).
\label{eq:oHL} 
\end{align}
\end{subequations}
The last terms in Eqs.~\eqref{eq:oHCT} and \eqref{eq:oHPT}
are the $A^2$ term and the transverse $P^2$ one, respectively.
The Hamiltonian \eqref{eq:oHL} for the longitudinal field
has the same form in both gauges,
and the last term is the longitudinal $P^2$ term.
Since the three orientations are orthogonal to each other,
the Hamiltonians $\oHC^{\vk,\xi}$ ($\oHP^{\vk,\xi}$)
are commutable with each other for different wavevector (except $-\vk$)
or orientation.
In this way,
when we discuss the small deviations from the normal ground state,
the transverse and longitudinal fields are not mixed
in the infinite, homogeneous, and isotropic systems,
and then the extraction of the Hamiltonians $\oHCO$ and $\oHPO$
along one direction of $\vk$ can be justified.


\end{document}